%% file: preprint.tex
\PassOptionsToPackage{dvipsnames}{xcolor}
\documentclass[10pt,sigconf,screen,authorversion]{acmart}

\usepackage[all]{nowidow}
\usepackage{xcolor}
\usepackage{subcaption}
\usepackage{hyperref}
\usepackage{acronym}
\usepackage{textcomp}

\usepackage[capitalise,noabbrev]{cleveref}
\crefformat{section}{\S#2#1#3} 
\crefformat{subsection}{\S#2#1#3}
\crefformat{subsubsection}{\S#2#1#3}

\begin{document}

\author{Tobias Pfandzelter}
\affiliation{%
    \institution{TU Berlin \& ECDF}
    \city{Berlin}
    \country{Germany}}
\email{tp@mcc.tu-berlin.de}

\author{David Bermbach}
\affiliation{%
    \institution{TU Berlin \& ECDF}
    \city{Berlin}
    \country{Germany}}
\email{db@mcc.tu-berlin.de}

\title{Can Orbital Servers Provide Mars-Wide Edge Computing?}

\acmYear{2023}\copyrightyear{2023}
\setcopyright{acmlicensed}
\acmConference[SatCom '23]{1st ACM MobiCom Workshop on Satellite Networking and Computing}{October 6, 2023}{Madrid, Spain}
\acmBooktitle{1st ACM MobiCom Workshop on Satellite Networking and Computing (SatCom '23), October 6, 2023, Madrid, Spain}
\acmPrice{15.00}
\acmDOI{10.1145/3570361.3614239}
\acmISBN{979-8-4007-0335-5/23/10}

\begin{abstract}
    Human landing, exploration and settlement on Mars will require local compute resources at the Mars edge.
    Landing such resources on Mars is an expensive endeavor.
    Instead, in this paper we lay out how concepts from low-Earth orbit edge computing may be applied to Mars edge computing.
    This could lower launching costs of compute resources for Mars while also providing Mars-wide networking and compute coverage.
    We propose a possible Mars compute constellation, discuss applications, analyze feasibility, and raise research questions for future work.
\end{abstract}

\maketitle

\input{sections/1_introduction.tex}
\input{sections/2_background.tex}
\input{sections/3_constellation.tex}
\input{sections/4_usecases.tex}
\input{sections/5_feasibility.tex}
\input{sections/6_conclusion.tex}

\begin{acks}
    Funded by the \grantsponsor{BMBF}{Bundesministerium für Bildung und Forschung (BMBF, German Federal Ministry of Education and Research)}{https://www.bmbf.de/bmbf/en} -- \grantnum{BMBF}{16KISK183}.
\end{acks}

\balance

\bibliographystyle{ACM-Reference-Format}
\bibliography{bibliography.bib}

\end{document}

%% file: sections/1_introduction.tex
\section{Introduction}
\label{sec:introduction}

Mars is the next frontier in human exploration and settlement~\cite{farley2020mars,drake2010human}.
Scientific probes, autonomous rovers, and human habitats on Martian soil will require a range of network and compute infrastructure for life-support systems, communication services, and scientific data analysis with machine learning (ML)~\cite{edwards2006relay,radanliev2020design,wright2023lessons}.
The six-minute round-trip time (RTT) for signals between Mars and Earth~\cite{edwards2006relay} is too large to run all of these systems on Earth computers -- instead, Mars will require its own ``Mars cloud''.

Until humans can efficiently harvest Mars' natural resources and build sophisticated data centers from them on the Mars surface, computers and networking components will need to be flown in from Earth.
Crucially, this not only requires launching them on rockets from Earth, but also equipping them with landing mechanisms for Mars, which increases weight, cost, and mission risk~\cite{farley2020mars,spaceexploration61}.

On Earth, large low-Earth orbit (LEO) satellite constellations are increasingly used to provide global broadband Internet access~\cite{Handley2018-ay,Bhattacherjee2018-vc}.
Studies have also shown the feasibility of equipping such satellites with compute resources to provide in-network computing services in LEO~\cite{Bhattacherjee2020-kr,Bhosale2020-aa,techreport_pfandzelter2022_celestial_extended,paper_pfandzelter2022_celestial,wang2021tiansuan,paper_pfandzelter2022_LEO_placement}.
Compared to terrestrial Earth data centers, such compute satellites have high costs, are complex, and provide only constrained resources that make them useful only for niche use-cases.

On Mars, however, two characteristics could make compute satellite constellations a viable option:
First, providing compute services from orbit would obviate the need for Mars landing equipment, reducing cost and risk.
Second, orbital characteristics mean that a satellite constellation with only tens of satellites could provide Mars-wide coverage, increasing flexibility for missions and human settlement.

In this paper, we provide a preliminary analysis of the feasibility of this proposal.
We make the following contributions:

\begin{itemize}
    \item We propose a small, 81-satellite low orbit constellation for Mars compute and communication services (\cref{sec:constellation}).
    \item We discuss possible applications for such a constellation (\cref{sec:usecases}).
    \item We provide a first analysis of the feasibility of providing compute and networking from Mars orbit (\cref{sec:feasibility}).
\end{itemize}

%% file: sections/2_background.tex
\section{Background}
\label{sec:background}

\subsubsection*{Mars}

Mars is the fourth planet from the Sun, close to Earth, and the most similar to Earth from any planet in the solar system.
Mars is a terrestrial planet with a mean radius of 3,389.5km (53.2\% Earth radius) and rotates once every 24.6 hours~\cite{marsfactsheet}.
Mars has little atmospheric pressure, only about 1\% as dense as that of Earth at sea level~\cite{marsweather}.
Nevertheless, weather effects in the form of dust storms are common on Mars, with large annual storms that cover continent-sized areas for days or weeks and \emph{global dust storms} that cover the entire planet an average of once every 5.5 Earth years~\cite{marsstorms}.

\subsubsection*{Mars Exploration \& Occupation}

To date, Mars has been explored exclusively remotely by spacecraft such as rovers, probes, and helicopter.
Although human exploration and settlement of Mars has been proposed and discussed as early as the 1950s~\cite{dasmarsprojekt}, more recent concepts and proposals by governmental space agencies and private aerospace companies have targeted launch years in the 2030s~\cite{drake2010human,wooster2018spacex}.
Missions are usually planned to coincide with Mars launch periods, roughly every 26 Earth months, where energy required to transfer between Earth and Mars orbits are lowest~\cite{marslaunchwindows}.
There are many potential sites for human landing and settlement on Mars, dictated by availability of ice deposits, risk of dust storms, and temperature.
Current analyses suggest that the equatorial region of Mars fits these criteria best~\cite{marslandingsite}.

\subsubsection*{Mars Relay Network}

Data transfer between Mars and Earth, e.g., to support the \emph{Perseverance} rover, is handled by the \emph{Mars Relay Network}, piggybacking onto NASA and ESA Mars orbiters~\cite{9843762,edwards2006relay}.
This limits the required mass for Mars landers, as they only need power and antenna for radio communication to Mars orbit rather than direct-to-Earth communication.
As of now, this network provides only connections between Mars and Earth (not between multiple parties on Mars) and is only used for applications with  limited bandwidth (40.5kbit/s)~\cite{9843762}.

\subsubsection*{LEO Satellite Networks}

Public and commercial actors are building communication constellations comprising hundreds or thousands of satellites in LEO~\cite{Handley2018-ay,Bhattacherjee2018-vc}.
Traditional satellite Internet relays were deployed in geostationary orbit at altitudes of more than 35,000km, inducing communication delays of 550ms RTT~\cite{10.1145/3517745.3561432}.
Technological advances have made LEO satellites at altitudes of less 2000km possible.
The smaller cone of coverage for each satellite necessitates i)~more satellites per constellation to provide global coverage and ii)~optical inter-satellite links (ISL) between satellites to connect distant ground stations with high bandwidth without terrestrial relays~\cite{Handley2018-ay,Bhattacherjee2018-vc}.
Further, orbital mechanics mean that satellites travel at high speeds in relation to Earth, e.g., 27,000km/h at 550km altitude~\cite{Bhattacherjee2019-jz}.
As a result, ground stations frequently connect to different satellites.

\subsubsection*{LEO Edge Computing}

Researchers have proposed extending LEO satellite networks with compute resources to provide orbital in-network computing~\cite{Bhattacherjee2020-kr,wang2021tiansuan}.
Similarly to terrestrial edge computing, this could provide low latency compute services to clients and reduce network strain for bandwidth-intensive applications, including metaverses or the Internet of Things.
The key challenges of LEO edge computing are deploying application services on limited resources and counteracting the highly dynamic LEO satellite movement~\cite{paper_pfandzelter2021_LEO_serverless}.

%% file: sections/3_constellation.tex
\section{Low-Mars Orbit Constellation}
\label{sec:constellation}

The Mars equivalent to geostationary orbit is the \emph{areostationary} orbit about 17,000km above Mars' surface~\cite{lay2001developing}.
Assuming a 25° minimum angle of elevation for ground station equipment~\cite{Bhattacherjee2019-jz}, four areostationary satellites are enough to cover the entire equatorial circumference of Mars.
At this altitude, however, only ground stations below the 56.3° latitudes can access the network.
While this may be enough for early human settlements, it excludes much of the planet, including its polar region.
Even more important, a high altitude also leads to higher transmit power requirements:
Lay et al.~\cite{lay2001developing} calculate a 10W requirement for a reliable 1kb/s link from the ground to this altitude, compared to on the order of 100mW for a link to 1,000km orbits.
Communication RTT with a ground station on the surface is also reduced by 90\% from 125ms (areostationary orbit) to 12ms (1,000km orbit).

\begin{figure}
    \centering
    \includegraphics[width=0.9\linewidth]{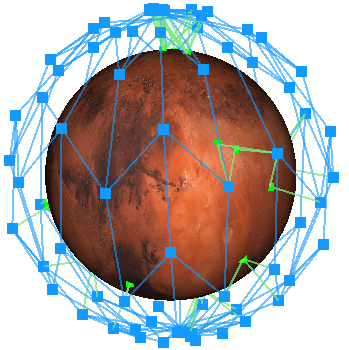}
    \caption{A Walker Star satellite constellation around Mars with 81 satellites, with nine orbital planes of nine satellites each at 1,120km altitude. Green points show ground stations at Mars probe landing sites~\cite{marslandingsites}.}
    \label{fig:constellation}
\end{figure}

The downside of lower orbits is the number of satellites required for Mars-wide coverage.
We show a possible constellation for Mars-wide communication coverage in \cref{fig:constellation}~\cite{9197767}\footnote{We make this simulation tool available as open-source: \url{https://github.com/pfandzelter/mars-orbit-simulator}.}.
This is a \emph{Walker Star} constellation (similar to the Iridium constellation~\cite{408677}) of nine orbital planes with nine satellites each, a total of 81 satellites, each at an altitude of 1,120km.
We assume optical ISL in a \emph{+GRID} topology~\cite{Bhattacherjee2019-jz}.
By adding compute resources to this satellite constellation, it could provide Mars-wide edge computing capabilities.
The maximum RTT from a ground station to a satellite server is 12.5ms, which is sufficient for most applications~\cite{Mohan2020-cn}.
While this may not be the perfect constellation design for Mars, depending on specific network and quality-of-service (QoS) requirements~\cite{fugmann2020automated}, it is a useful starting point to discuss possible network characteristics of a low-Mars orbit communication and compute constellation.

%% file: sections/4_usecases.tex
\section{Applications}
\label{sec:usecases}

We envision a range of applications that could benefit from such a compute constellation.

\subsection{Computation Offloading}

\begin{figure}
    \centering
    \includegraphics[width=\linewidth]{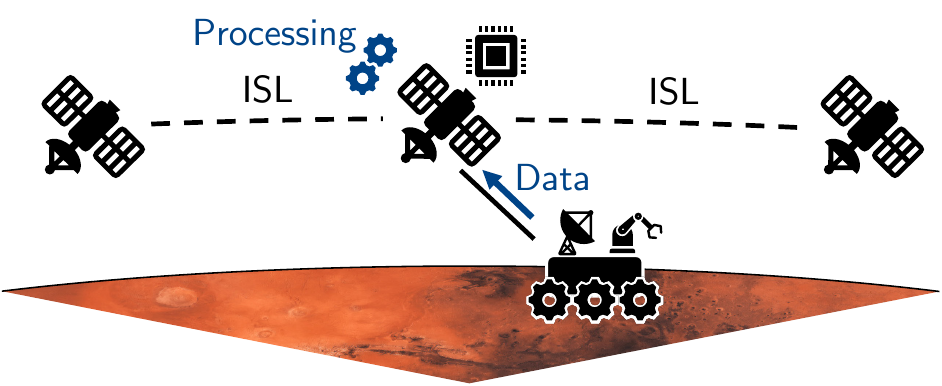}
    \caption{A device on Mars could offload intensive data processing, e.g., ML inference, to the compute constellation to reduce energy consumption.}
    \label{fig:offloading}
\end{figure}

An obvious use-case for a Mars compute constellation is offloading computationally intensive tasks from low-power devices, such as autonomous rovers, scientific instruments, or sensor networks.
We show an example for such an application in \cref{fig:offloading}.
The rover in this example generates data by sensing and interacting with its environment.
Processing of this data, e.g., using ML inference, is resource intensive.
Instead of equipping the rover itself with the necessary compute resources, cooling system, and power source, it could efficiently offload processing to the compute constellation with low latency.

\subsection{Multi-Party Collaboration}

\begin{figure}
    \centering
    \includegraphics[width=\linewidth]{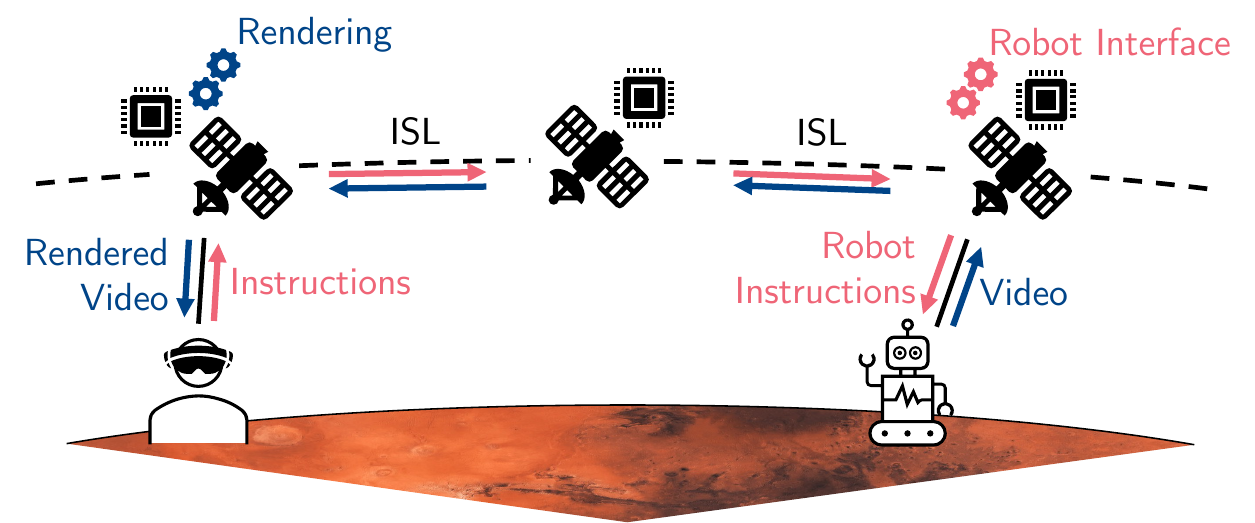}
    \caption{A collaborative application where a user remotely controls a Mars robot. Video rendering and robot interface services can run within the network path between the parties, adding little latency to the interaction.}
    \label{fig:collaboration}
\end{figure}

Embedding compute resources in the network adds the benefit of low latency processing on the network path to offloading.
Applications where multiple parties interact, e.g., different users or devices, can benefit from low additional network delay when offloading processing tasks.
An example for such a collaborative application is shown in \cref{fig:collaboration}.
The user on Mars remotely controls a robot with an immersive video interface.
The satellite network provides low-latency interaction between the two parties, and the required services for rendering video and interfacing with the robot can be deployed within the network path, without additional communication delay~\cite{Bhattacherjee2020-kr,Mohan2020-cn}.

\subsection{Caching Incoming Data}

\begin{figure}
    \centering
    \includegraphics[width=\linewidth]{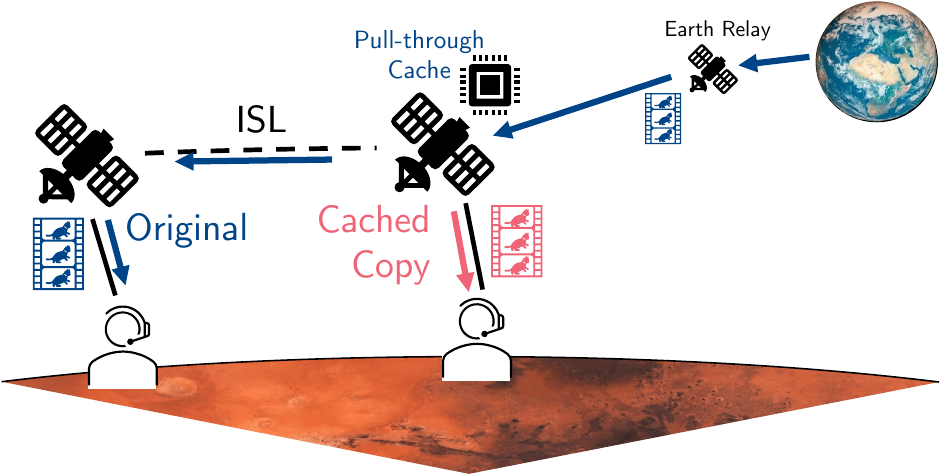}
    \caption{A pull-through cache reduces strain on the limited bandwidth link between Mars and Earth: by caching frequently requested files directly in the uplink network, users can benefit from significantly lower access latency.}
    \label{fig:caching}
\end{figure}

Similarly to a content delivery network (CDN), the compute constellation could also support caching incoming data from Earth~\cite{paper_pfandzelter2021_LEO_CDN,Bhosale2020-aa}.
Instead of every client requesting a specific file over the low-bandwidth and high-delay link to Earth, the file could be cached after the first request (or even pushed from the origin location if requests can be anticipated).
As illustrated in \cref{fig:caching}, subsequent requests for this file can be served from a local cache, reducing access latency.

\subsection{Pre-Processing Outgoing Data}

\begin{figure}
    \centering
    \includegraphics[width=\linewidth]{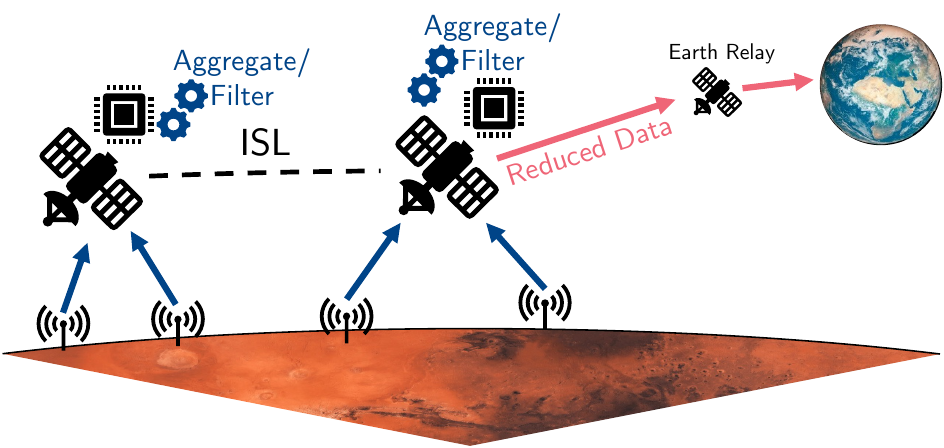}
    \caption{Preprocessing data on the compute constellation using aggregation or filtering reduces the required uplink bandwidth for data sent to Earth without increasing resource requirements of the sensors.}
    \label{fig:preprocessing}
\end{figure}

Deploying compute services on the path between may also be used to reduce the amount of data relayed back to Earth:
The example in \cref{fig:preprocessing} shows a Mars sensor network sending data to the satellite network in order to relay it back to Earth.
By deploying aggregation and filtering services to the satellite servers, data volume can be reduced before it is sent to Earth~\cite{paper_bermbach2017_fog_vision}.

%% file: sections/5_feasibility.tex
\section{Feasibility}
\label{sec:feasibility}

The feasibility of edge computing in satellite networks has been shown for Earth~\cite{Bhattacherjee2020-kr,pfandzelter2023failure}.
While the general architecture is similar for a Mars constellation, we identify three environmental differences that could impact the feasibility of satellite edge computing for Mars.

\subsection{Radiation Environment}

The radiation environment in LEO is subject to Earth's magnetic field and the Van-Allen radiation belts~\cite{nasa2014vanallen,pfandzelter2023failure}.
Little aluminum shielding is required to protect commercial off-the-shelf compute components for a five-year satellite lifetime~\cite{maurer2008td}.
With this shielding, a processor has an expected soft error rate on the order of $10^{-3}$ to $10^{-4}$ per day~\cite{pfandzelter2023failure}.

\begin{figure}
    \centering
    \includegraphics[width=\linewidth]{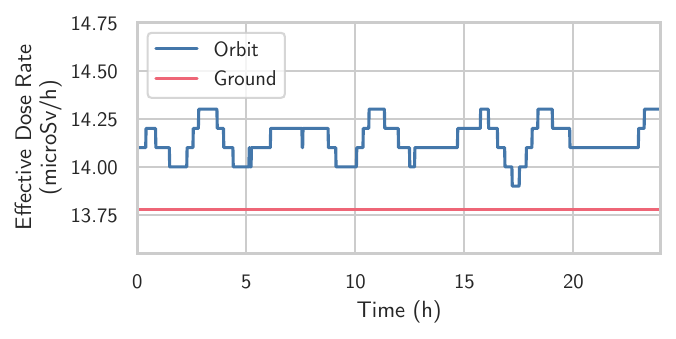}
    \caption{Radiation environments in 1,120km orbit and on Mars ground compared by means of the effective dose rate as calculated by MEREM~\cite{gonccalves2009marsrem}.}
    \label{fig:radiation}
\end{figure}

Mars lacks a magnetic field, exposing objects in orbit to galactic cosmic rays and solar particles~\cite{semkova2023observation}.
Yet this also affects equipment on the Mars surface, as \cref{fig:radiation} shows:
A calculation using the \emph{Mars Energetic Radiation Environment Models (MEREM)}~\cite{gonccalves2009marsrem} shows that the effective dose rate is comparable between Mars surface and an 1,120km altitude orbit.
The effective dose rate is a measure of particle flux as it affects the human body~\cite{valentin20072007} and gives an impression of radiation levels, although it is not directly applicable to electronic equipment.

We leave a more detailed comparison of Earth and Mars radiation levels and its constraints for satellite server hardware for future work.
We note, however, that also radiation effects during the transportation of components from Earth to Mars must be taken into account.

\subsection{Impact of Mars Weather}

On Earth, heavy rain can impact the up- and downlink performance of satellite-based Internet access, decreasing throughput by 50\%~\cite{ma2022network,kassem2022browser}.
Mars dust storms could degrade high-bandwidth radio links similarly, but their effect is likely not as large as that of Earth storms:
Ho et al.~\cite{ho2002radio} find that despite their size, large Mars dust storms attenuate K\textsubscript{a}-band radio links only 3dB in the worst case as a result of the small size of Mars dust particles (1-4\textmu{}m).
This is comparable to K\textsubscript{a}-band link attenuation observed by Vasisht et al.~\cite{vasisht2021l2d2} with cloud cover.

\subsection{Cost Reduction}

Sending any equipment or even humans to Mars is also a costly endeavor:
To give an example, NASA's Perseverance cost an estimated 2.2 billion USD to develop~\cite{perseverancecost}.
Launch and development costs have to be reduced by orders of magnitude before human settlement and large-scale satellite constellations on Mars can even be considered.

Nevertheless, we can already roughly estimate the cost benefits of providing Mars compute services from orbit instead of from the ground:
Assuming no existing power and cooling infrastructure on Mars, each compute resource built on Earth and launched to Mars must carry its own power generator and cooling equipment.
When such a server is transferred to Mars, it is first inserted to Mars orbit.
Hence, the cost delta between satellite and ground servers is predominantly the landing equipment necessary to land on Mars safely.
This includes heat shielding to protect hardware from entry into the Mars atmosphere, parachutes, retro rockets, landing legs, and airbags.
Consider as an example the \emph{Mars 2020} mission that included the Perseverance rover:
The rover itself weights 1,025kg, comparable to the existing 800kg Starlink V2 satellites that include solar arrays, batteries, Hall-effect thrusters, and radio antennas but no high-performance compute servers~\cite{starlinkv2mini,Selva2017412}.
The entry, descent, and landing (EDL) architecture required for Mars 2020, however, comprises a 575kg backshell, 440kg heat shield, and 670kg descent stage with 400kg propellant for a total of 2,085kg~\cite{spaceexploration61,farley2020mars,nelessen2019mars}.

While this is just a preliminary calculation, the more than 200\% mass overhead for landing equipment shows that providing networking and compute services from orbit can be a significant opportunity for cost savings.

%% file: sections/6_conclusion.tex
\section{Conclusion \& Future Work}
\label{sec:conclusion}

Exploration of and human settlement on Mars is still many decades away, yet we argue that discussing the possibilities of networking and computing services that may be provided on Mars is already relevant today.
Importantly, our preliminary evaluation of orbital compute services for Mars show that existing research in LEO networking and edge computing is not just relevant for niche use-cases on Earth but may also become relevant for other planets in our solar system.

Of course, there are still many open questions that must be addressed, including if human occupation of Mars is desirable at all.
Among those, hardware design is of key concern, as weather, radiation, and atmosphere on Mars are different from those known on Earth.
Further, this research may also extend to other planets and celestial bodies that humans may explore further in the future, e.g., Moon or Venus.